\documentclass[aps,prl,twocolumn,superscriptaddress,nofootinbib,
nobalancelastpage,nobibnotes]{revtex4}
\usepackage{epsfig}
\usepackage{graphicx}
\usepackage{caption}

\begin{document}

\title
{Quantum breaks in a model for the evolution of neutrinos during their decoupling era in the big bang}

\author{R. F. Sawyer}
\affiliation{Department of Physics, University of California at
Santa Barbara, Santa Barbara, California 93106}

\begin{abstract}
An active two-neutrino state's coupling to an intermediate scalar meson, and thence to a two-antineutrino state, is one idea for catalyzing transitions from active to sterile neutrinos, in an implementation of the Dodelson-Widrow proposal for the production of sterile and somewhat massive neutrinos as dark matter candidates. 
We propose some mechanics that promises to use the very same model with the same mass and coupling parameters, in order to achieve similar results, but orders of magnitude faster; or alternatively, similar results but beginning with vastly reduced coupling constants. The model then would become much less constrained by consistency with laboratory and astrophysical data.

\end{abstract}
\maketitle
\section{1. Introduction}
Recently the reaction $\bar \nu_e+\bar \nu_e  \leftrightarrow \nu_x+\nu_x $, as mediated by an intermediate scalar meson, in conjunction with a standard bilinear coupling with a sterile neutrino
$\nu_s$, has been used \cite{deg} in a promising implementation of the Dodelson-Widrow proposal 
\cite{DW} for the production of dark matter as composed of sterile neutrinos.
Here $\nu_x$ stands for another active neutrino species, $\nu_\mu$ or  $\nu_\tau$. The time frame 
is centered in the era of conventional neutrino decoupling.
 
Another recent work \cite{ful} has addressed the changes in standard early universe results in the same scalar coupling model, but leaving out the sterile neutrino possibility altogether. These authors find some relatively minor changes but ones that might be checked in the current era of ever more precise data. The present paper bears on both of the above-cited works. 

In both exercises we have $\bar \nu_e+\bar \nu_e  \leftrightarrow \nu_x+\nu_x $, in which all participating neutrinos are left-handed,  coupled through an intermediate scalar meson that has no other couplings except to active neutrinos. Here $\nu_x $ stands for some mixture of $\nu_\mu$ and $\nu_\tau$. In both of the works cited above, this reaction enters only through its cross-section
in vacuum. This is of the order $G^2$ in some generic 4-Fermi coupling constant $G$, leading to a characteristic time $T_c$ for a macroscopic effect of $T_c  \sim [G^2 E_\nu^2 n_\nu]^{-1}$. Our replacement, based on coherent many-body amplitudes, is capable, under some circumstances, of achieving total flavor mixing in time   $T_c  \sim [G n_\nu]^{-1} \log \Lambda$, where $\log \Lambda$ could be as big as 30. The word ``rate" is inapplicable here since the behavior consists of a long wait with only a tiny bit happening, followed by a rapid transition.

We shall consider only time intervals in which ordinary scattering in the medium is negligible and for which $T <<m_\nu^{-2} E$, in order to preserve the coherence that drives the effect. And we consider only amplitudes in which, in any two-$\nu$ sub-amplitude, only reactions like 
${\bf \nu(p)+\nu(q) \leftrightarrow \bar \nu( p)+\bar \nu( q)}$ enter, with momentum
``preservation", not just momentum conservation. With genuinely massless particles (such as the gravitons and photons considered in ref. \cite{rfs1}) we then would have energy and momentum exactly conserved in every reaction.

Fig. 1 illustrates a miscellaneous 8-$\nu$ process, showing some conversion vertices (black-circles) amidst a swarm of eight neutrinos that are incoming from different directions. It is amazing at first glance that there is a coherent part to this graph that would survive averaging over angles when we are simulating an initial isotropic distribution and then summing over energy distributions. The reader can imagine how the combinatorial factors build up in a complete perturbative expansion with the numbers of particles that participate coherently increasing, order by order, as a power of total particle number N.  In what follows we develop the applicable equations for the non-perturbative approach and solve them numerically. We do all calculations in a periodic box somewhat larger than $cT$.  

 \begin{figure}[h] 
 \centering
\includegraphics[width=2.5 in]{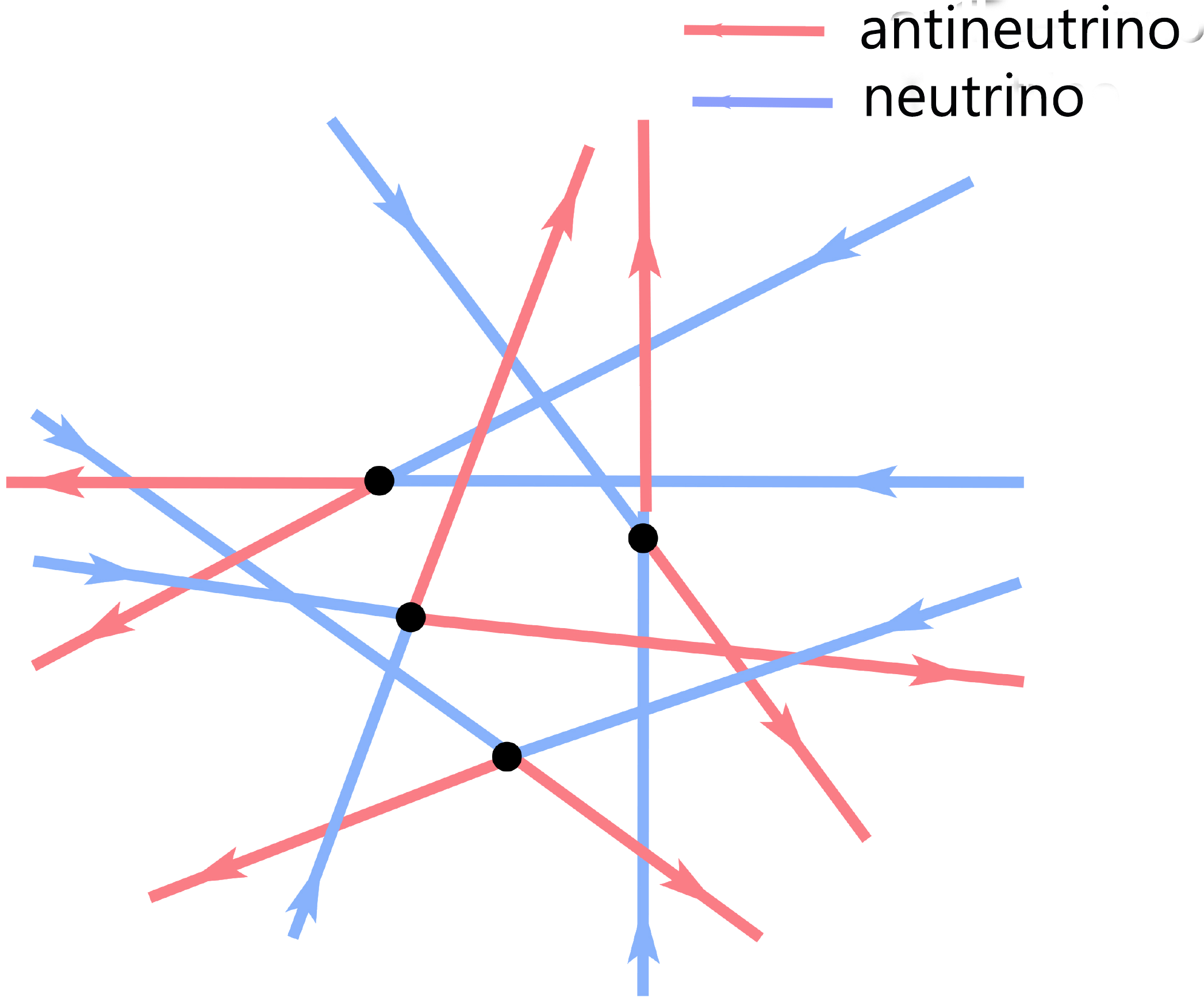}
 \caption{ \small } 
 \label{fig. 1}
\end{figure}
Secs. 2-5 are devoted to the system without the coupling to sterile $\nu$'s.
Sec. 6 incorporates the mixing with them, but is less definitive, because of computational stresses connected to getting to the extremely small mixing angles that the literature demands. 
\section{2. Analytic treatment of models.}  By now several types of systems with features in common with the ones
we shall discuss have been analyzed: a)  Bose condensates of atoms in wells,\cite{va}-\cite{va3};
b) polarization exchanges in collisions of high intensity photon beams \cite{rfs1};  c)``Fast" neutrino flavor transformation at the supernova neutrino-sphere \cite{rfs3}-\cite{f9};  d) A cloud of very long wavelength gravitons at very high number density (in vacuum otherwise) transforming to photon pairs in a time proportional to $G^{-1}$, \underline{not} as $G^{-2}$ \cite{rfs6}.

Based on this experience,
we state requirements for our present work:
 \subsection{A. Mean field} We require there to be a sensible mean-field theory of the phenomena.  The standard formulation as given in detail by Raffelt and Sigl \cite{rs} is the background for everything we do. But it will be modified in what is to follow, in which the vector four-Fermi coupling is replaced by a scalar coupling. Then in $\nu+\nu \leftrightarrow \bar \nu + \bar \nu $ the reactions proceed with all of the $\nu,\bar\nu$'s staying within the left-handed sector. The basic variables are
$ \bar\nu^c_k \nu_j$ and $ \bar\nu_k \nu^c_j$, quadratics in the neutrino fields, with $j,k$ as flavor indices.
The notation $\nu^c_j$ stands for the charge-conjugate field.
\subsection{B.  Instability}
Within the standard mean field context, if we begin with a pure unmixed flavor state, it is clear that nothing whatever would happen. But once the mass (i.e. $\nu$ oscillation) terms begin to mix states, then a lot  \underline{may} happen in a very short time, at least in the much-studied vector-coupled case.
Analytically we can distinguish such cases by linearizing the non-linear equations of evolution, through taking the diagonal matrices in flavor space as fixed at their initial values. In the usual neutrino case we then may find exponentially growing modes signaled by the real parts of the eigenvalues of a simple matrix. In the usual vector coupling case, the driving instability exists only in the case of certain initial angular distributions and spectra. In contrast, in the 
$\nu+\nu \leftrightarrow \bar \nu+\bar \nu$ case the driving instability will be present for any angular distribution, at least in the case in which we have near perfect particle-antiparticle symmetry of the medium. But if we start with a flavor-diagonal initial state the system 
needs a tiny seed from somewhere in order to exploit the instability.

\subsection{C. Quantum break} Taking the neutrino masses as the seed for the instability in all of the ``fast transformation" literature for the standard vector coupled case obscured a fascinating fact, namely that the same fast processes for the unstable cases would have turned themselves on anyway, even in the absence of neutrino oscillations, and on nearly the same time scale, through a  ``quantum break". 
That term was used, for example, in a classic article on Bose condensates of atoms in wells \cite{va}. The question then addressed was: when, in the context of mean-field theory, we find a system
that is in unstable equilibrium, how do we formulate the theory of the fluctuation that gives it a push to get it moving? Our test system of quantum fields spread out over a macroscopic region that contains 10$^{30}$ particles is far different from 100 atoms in a well, but the methods that can be applied are related. 

In our new application of similar ideas to the scalar interaction $\nu+\nu \leftrightarrow \bar \nu+\bar \nu$ the quantum break becomes an indispensable necessity, because we are not introducing a lepton-number breaking interaction that mixes $\nu$ and $\bar \nu$ on a single-particle basis, which otherwise could have provided a seed for the instability. We shall develop a modified form of mean-field theory that provides a solution that withstands testing in soluble systems of $N \le 1000$ neutrinos, over the time scale that we need to consider. Then, with fingers crossed, and some consistency checks, we apply it to systems with $10^{30}$ $\nu$'s.
  
 \section{3. Solutions} 
We are working toward describing the development of systems containing $e^{\pm}$, 
$\nu, \bar \nu $, and photons, and which embody perfect particle-antiparticle symmetry and isotropy. But we consider only time intervals that are short on the scale of free paths for scattering, and in which the neutrino-neutrino interactions, if unstable, can dominate the dynamics. And in calculating coherent effects we must
trace what each quantum state of the multi-particle system does, before adding up to see how the statistical ensemble (which is where the isotropy, e.g., is encoded) evolves.

We begin by considering a beam of $\nu_e$'s in a narrow angular cluster impinging on another beam of $\nu_e$'s to make reactions,   $2\nu_e \leftrightarrow 2 \bar \nu_x$, where $\nu_x$ is another active flavor.
Let $a^j$ be the annihilator for $\nu_e$ with momentum $\vec p_j$ in one incident 
$\nu_e$ beam and $b^k$ the annihilator for a state $\vec q_k$ in the opposed $\nu_e$ beam; with
$c^j , d^k$ being the annihilators for the same two momentum states of the $\bar \nu_x$ system. Now we wish to take the piece of the four-Fermi interaction that is the local limit  for the processes $ \nu_e+ \nu_e  \leftrightarrow \bar\nu_x+\bar \nu_x $ when the intermediate scalar meson mass is large compared to the neutrino energies (eq. 1.1 of  \cite{ful} )  and express the answer in terms of the operators $a,b,c,d$. 

We define the bilinears $\sigma_+^j =c_j^\dagger a_j $ for  $j=1$ to $N$,  and
$\tau_+^k =d_k^\dagger b_k   $, for $k=1$ to $N$, together with their Hermitian conjugates, 
$\sigma_-^j $, $\tau_-^k$.  We shall also use $\vec \sigma^{\, j}\,,\vec \tau^{\,k}$ as sets of 3-vectors that separately have the commutation relations of independent Pauli spin matrices, e.g. $ [\sigma_+^j,\sigma_-^k]=\delta^{j,k} \sigma_3^j$ etc.

From these we construct an effective Hamiltonian that embodies the action of the coupling through the scalar meson intermediary as long as the mass of the scalar is large compared to the neutrino energies,
\begin{eqnarray}
H_{\rm eff} = { 2\pi G  \over   V} \sum_{j,k}^N \Bigr [ \sigma_+^j  \tau_+^k+ \sigma_-^j \tau_-^k\Bigr ] (1-\cos \theta_{j,k})~.
\label{ham1}
 \end{eqnarray}
 
The above form embodies terms from the interaction Lagrangian
$ -\mathcal{L_I }$ of eqn. 1.1 in \cite{ful}, but simplied by taking all flavor-diagonal couplings of the
scalar field to vanish, and coupling only to a single pair of neutrinos, $\nu_e, \nu_x$. This produces 
not only the terms in (\ref{ham1}) but some additional additional terms that do not contribute to results in the large N limit.

However, for the sake of transparency in this first look at this category of model, we change the framework slightly to rid ourselves of these terms, which are unwanted complications at their least. We take a complex scalar field, $\Phi$ to mediate the interaction, massive on the scale of the energies, with coupling,

\begin{eqnarray}
\mathcal{L}_{\rm int}=g_{i,j} \bar\nu_{i}^c\,\nu_{j,l} \Phi+g_{i,j} \bar\nu_{i}\,\nu_{j} ^c\Phi^*
\label{lint}
\end{eqnarray}
exactly as in ref. \cite{ful}, eq. 1.1, and then specializing to the case $g_{1,2}=g_{2,1}=\sqrt G$ , 
$g_{1,1}=g_{2,2}=0$. Now (\ref{ham1}) needs no supplementation. Note that in (\ref{lint})
$i$ and $j$ are flavor indices while in (\ref{ham1}) they are momentum labels.

Because we are extracting only the part in which momentum
flows as $\vec p+\vec q \rightarrow \vec p+\vec q$ for particles that are nearly massless,
kinetic energy is conserved and the kinetic term in the Lagrangian is irrelevant. The answer above is independent of the absolute values of momentum that enter. But the directions (in whatever frame we choose for the calculation) enter only through the final factor in (\ref{ham1}) where 
	$\cos \theta_{j,k} =\vec p_j \cdot \vec p_k /(|p_j| |p_k |) $.

Our plan is first to solve the model numerically, but precisely, for the simplest case of head-on collision of two beams; this for the largest number of particles our modest computer resources can manage. 
Next we shall test the mean-field (MF) approaches (which are to come) that will be applied to cases with orders of magnitude more particle number. Finally we shall shift from two clashing beams to spherically symmetric initial distributions. 

For the head-on case we replace the angular factor in (\ref{ham1}) by 2. Then we sum over a set of momentum magnitudes indexed by $j,k$ to define collective operators $\vec \sigma=N^{-1/2}\sum_j^N \vec \sigma^{j}$ etc, and a Hamiltonian 
\begin{eqnarray}
H_{\rm eff} = {4 \pi  n G}[\sigma_+ \tau_+ + \sigma_- \tau_-]~,
\label{ham2}
\end{eqnarray}
where $n=N/V$ is the neutrino number density. 
The sets $\sigma_\pm, \sigma_3$,  $\tau_\pm, \tau_3$ still have the commutation rules of Pauli matrices; that is of angular momentum matrices times two. With the same number $N$ of $\nu_e$'s initially moving in the +$\hat z$ direction and of  $\bar \nu_e$'s in the $-\hat z$ directions, there is a single ladder of states connected by the interaction (\ref{ham1}). Taking $k$ as the index for the number of $\nu_x$ pairs that
have been created from our initial state with a prescribed set of momenta we have the matrix elements
\begin{eqnarray}
\langle k+1 |H_{\rm eff}|k \rangle=g k (N-k+1),
\label{schrod}
 \end{eqnarray}
for $k=1,...N+1$, where $g=  4\pi G   V^{-1} $. For small values of $N$ we can easily solve the Schrodinger equation based on (\ref{schrod}), numerically, to find the time dependent wave function of the system. We define the retention amplitude,
\begin{eqnarray}
 \zeta(t)=N^{-1}\langle \Psi(t)| (1/2+\sigma_3/2)|  \Psi(t)\rangle  \,,
 \label{zeta}
 \end{eqnarray}

 In fig. 2 we plot results for $\zeta (s)$ for a series of three N's spaced by powers of 4 from N=64 to 1024 as a function of scaled time $s=(g N)^{-1} t$.  As N is increased, the system's behavior will be to sit there for a longer time organizing itself, then making a sudden total turnover; that is as switch from particles of one flavor to anti-particles of the other. The actual turn-over time, in view of the scaling, is of the order of $(n G)^{-1}$, in contrast to a mean free time for scattering that is of order $(n G^2 E^2)^{-1}$, where $E$ is the energy scale of the cloud. 
 
Those remarks ignore the steady march to the right as N is increased in the curves shown in fig. 2; they are the hint of an additional $\log[N]$ dependence that we shall verify in the mean field approach.
  
In the ordinary mean-field approach the Heisenberg equations for the operators are easily constructed as, 
\begin{eqnarray}
\dot \sigma_+= n g \,\tau_-\, \sigma_3 ~~~~~~~,~~~~~~ \dot \tau_+= ng\, \sigma_-\,\tau_3\,,
\nonumber\\
\dot\sigma_3=-\dot \tau_3= 2 n  g(\sigma_+\, \tau_+ -\sigma_- \,\tau_-)\,.
\label{eoma}
\end{eqnarray}
The conventional mean field assumption (though the authors of ref.\cite{rs} do not phrase it quite in our terms) is that the operator equations (\ref{eoma}), are valid for the c-number expectations of the various operator products in the equations, $\langle A B\rangle = \langle A \rangle  \langle B \rangle$. 
But it is obvious that when we begin with pure flavor states, 
$\sigma_+(0)=0$, $\tau_+(0)=0$,  $\sigma_3(0)=-N $, $\tau_3(0)=N $, then nothing at all happens at the mean field level; the system stays exactly where it is. On the other hand if we
look just at the two coupled equations for $\sigma_+(t)$, $\tau_+(t)$ keeping $\tau_3=\sigma_3=N$
as fixed at the initial values, we see that a small perturbation, $\Delta \sigma_+$  will grow at a rate 
proportional to $\exp[nG t]$. The instability that is required for fast turnover is there. But the formalism has not supplied the break.    

\begin{figure}[h] 
 \centering
\includegraphics[width=2.5 in]{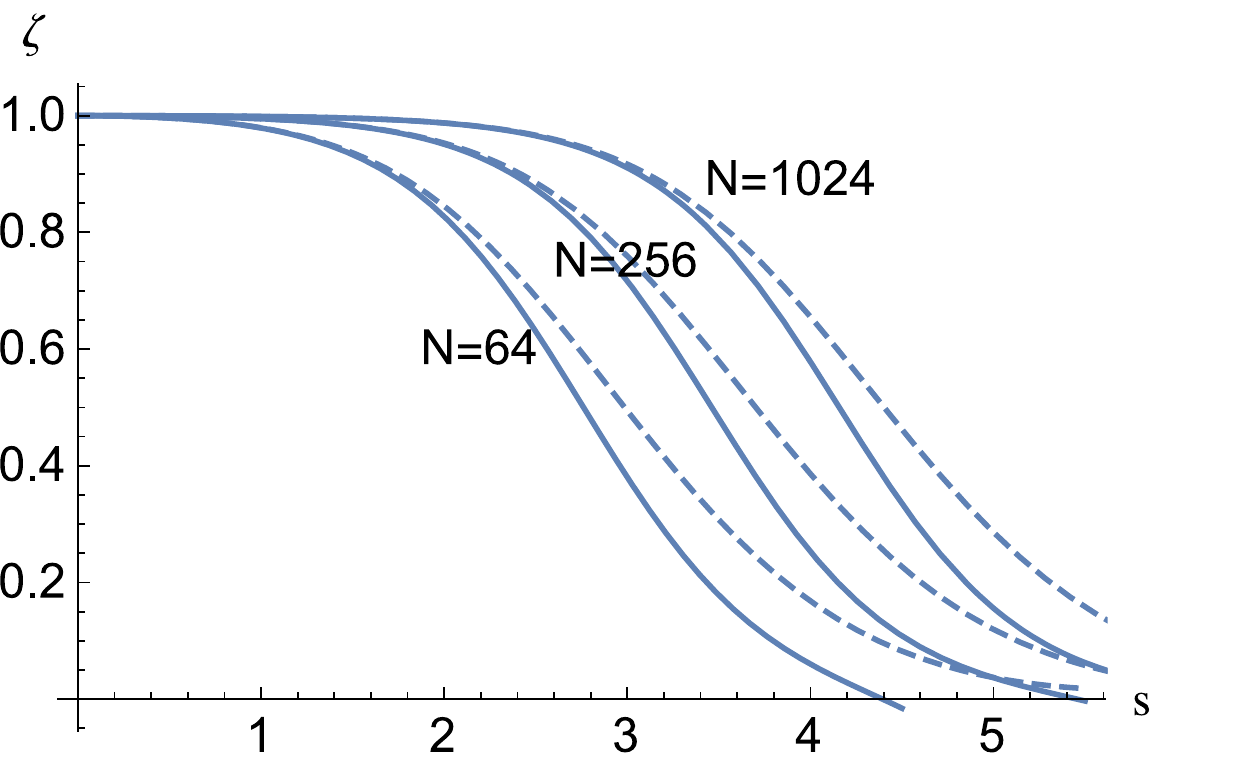}
 \caption{ \small }
Dashed curves: retention probability $\zeta (s) $in the wave function solution. Solid curves: the same in the corresponding MMF solutions, with their  $N$ values identifiable by their coalescence with the corresponding dashed curves for small scaled time $s$. $\zeta=0$ indicates 100\% flavor exchange between the two beams
\label{fig. 2}
\end{figure}

\section{4. Modified mean field approximation and quantum break}
The key to extending the MF approximation, as was done in the atomic physics problems of ref. \cite{va} and also 
for the collisions of circularly polarized protons with each other enabled through the Heisenberg-Euler coupling, in 
ref. \cite{rfs4}, is first to write equations of motion for some products of our basic operators. In our case we choose combinations
like $\sigma_+ \tau_+$. Now the right-hand-sides will contain higher order polynomials in the fields. We can then try
judiciously to close the system through factorizations of the generic forms 
$\langle ABC\rangle\rightarrow \langle AB\rangle\langle C\rangle$. 

The authors of ref. \cite{va} invoke references to BBGKY hierarchy in their description of their method, which is akin to ours, although the present author might say ``not even that", in the belief that it is much less than a quantum field theoretic form of what N. N. Bogolyubov had in mind \cite{bog}. 

Beginning with the operators defined in the last section $\vec \sigma$ for the ${\bf p}$ stream and  $\vec \tau$ for the ${\bf q}$ stream, and before rescaling,
we define $X=\sigma_+ \tau_+$, $Y=\sigma_+ \sigma_- + \tau_+ \tau_-$ . We rename  $\sigma_3= \tau_3=Z$ (their equality being chosen in the initial condition, and then maintained throughout).

The Hamiltonian is now
\begin{eqnarray}
H_{\rm g,\gamma}={8 \pi N G \over V}[X+X^\dagger] \,,
\label{mmfham}
\end{eqnarray}
and the Heisenberg equations of motion are,

\begin{eqnarray}
i \dot X ={8 \pi G\over V } (Z Y-Z^2) \, ,
\nonumber\\
i \dot Y={16 \pi  G \over V}  Z ( X^\dagger-X) \, ,
\nonumber\\
i \dot Z ={16 \pi  G \over V} (X-X^\dagger )\,.
\label{eom3}
\end{eqnarray}
The $Z^2$ term in the first equation comes from a second commutation to get operators into the correct order; implicitly it carries an additional power of $\hbar$ and is the source of the ``quantum break" to come.
Our modified mean field method (MMF) is to replace each of the operators $X, Y, Z$ in (\ref{eom3}) by its expectation value in the medium, thus implicitly assuming that, e.g, $\langle Z Y\rangle= \langle Z \rangle \langle Y\rangle$. 

Next we do a rescaling in which each one of the single particle operators $a, b, c, d, a^\dagger...$ is redefined by extracting a factor of $N^{1/2}$, so that $x=X/N^2$, $y=Y/N^2$, $z=Z /N$ and at the same time defining $n=N/V$, the number density, and the scaled time variable, $s$, according to \newline $s=8 \pi G n t$, where $n$=the number density
$N/V$ of each beam.
 
The rescaled equations are,
\begin{eqnarray}
i {d x \over ds} =zy-z^2/N \,,
\nonumber\\
i {d y \over ds} =2 z ( x^\dagger-x) \,,
\nonumber\\
i {d z \over ds} =2 (x-x^\dagger ) \,,
\label{mmf}
\end{eqnarray}

and the rescaled initial condition (all $\nu_e$'s at the beginning, say) is $z(s=0)=1$. The value $z(s)=-1$  signifies a complete transformation to $\nu_x$'s.
The retention fractions $\zeta(s)=(z(s)+1)/2$ calculated from the solutions of (\ref{mmf}) are plotted against scaled time, shown as the solid curves in fig. 2 for the same three values of $N$ used in that plot. We see an excellent fit at early times, in view of the fact that there are no free parameters, as the retention $\zeta(t)$ decreases 
from unity to .75 .  After that there is qualitative agreement down to the turn-around. There is also a hint that the fit is improving for higher values of $N$. 

We can also use the authoritative solutions of the preceding section to explore solutions of the ordinary mean field equations (\ref{eoma}) of the last section, where we have to provide some kind of seeding in order to make anything happen at all. We find that initial values of order $\sigma_+(s=0)\approx \tau_+(s=0)\approx \pm N^{-1} i$
give a point of $\zeta \approx 0$ at about the same time as found from the modified MF solutions for the three plotted cases, though not fitting the shape as well for the earlier times. 

In the modified mean-field approach we can extend the calculations to much higher vales of N, provided the beams have very small angular dispersion. Plots 
are shown in fig. 3.
\begin{figure}[h] 
 \centering
\includegraphics[width=2.5 in]{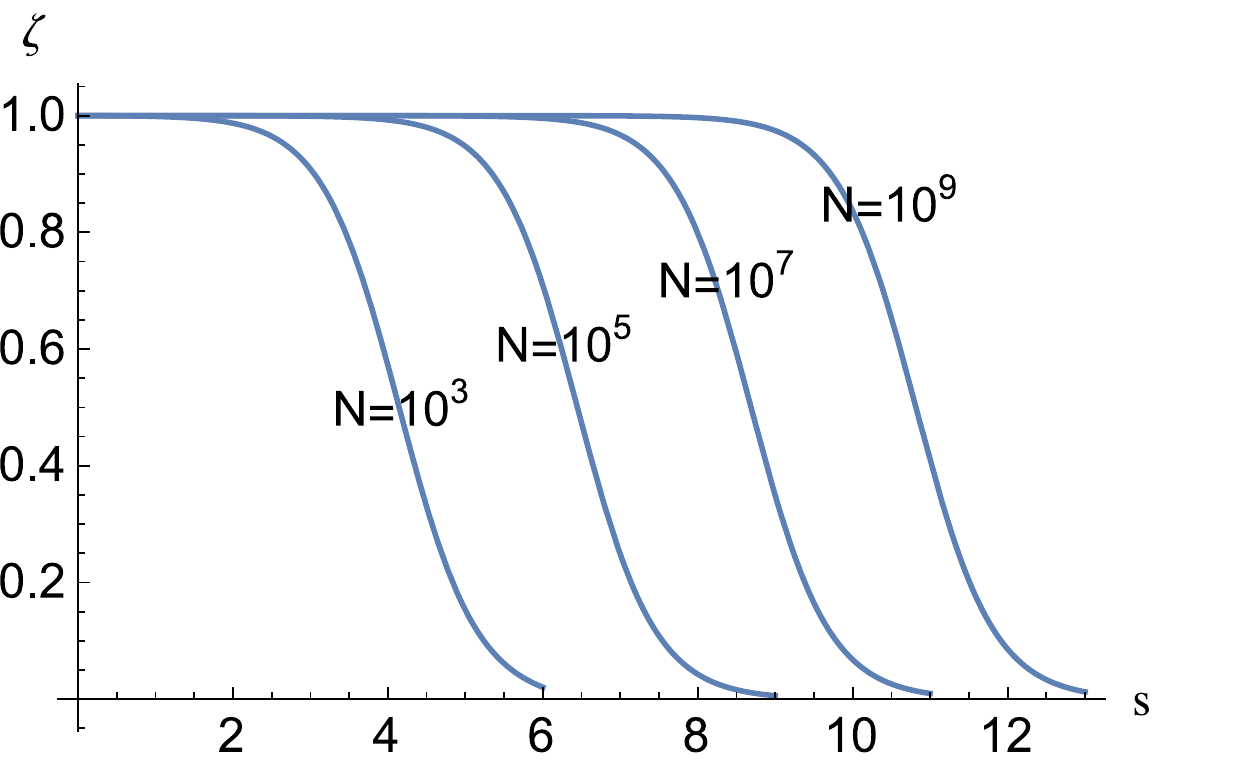}
 \caption{ \small }
The same as fig.2, except only the MMF solutions, for a series of higher values of N.
\label{fig. 3}
\end{figure}

\section{5. Multi-beam solutions, isotropy }

In the application to the early universe we seek the behavior for completely isotropic initial distributions
and in view of the $(1-\cos \theta)$ factor in the effective interaction, we might expect, at the least, some diminution of the effect. In addressing this situation we take a forest of angles, for $N_b$ different incident beams, each with $N_a$ neutrinos distributed uniformly in elements of solid angle $d \cos \theta \,d\phi$. This translates into 
an effective interaction,
\begin{eqnarray}
H_{\rm eff}={ 4 \pi G \over  V}\sum_{j,k}^{N_b} [\sigma^{j}_+ \sigma_+^{k} +\sigma_-^{j} \sigma_-^{k}] \lambda_{j,k} \,,
\nonumber\\
\label{hamxx}
\end{eqnarray}
where the $\lambda_{j,k}=(1-\cos \theta_{j,k}$) for the rays $j$ and $k$. 

The equations for the operators at the MF level involve the variables $\sigma^j_+,\tau^j_+$ 
and $\sigma^j_3= \tau^j_3$, and pose no particular problem in solution, but require
seedings of the form $\sigma^j_+ (0)= i (N_a N_b)^{-1}$. We have checked this latter claim by 
considering cases in which we try out combinations of $N$ factorized in different ways into $N_a N_b$ while maintaining the value $\lambda_{j,k}=1$ for all $N_b$ states, and observing that there is no difference in the results.

When we compare the head-on beams scenario with an
isotropic scenario that uses a 25 angle simulation in which each beam has the same fraction of the total solid angle we find that the results are the same except for one detail: in the isotropic case the time scale is greater by a factor of two. It is exactly as though $1-\cos \theta$ gets averaged over the sphere. Thus we have confidence that our effects will persist in an isotropic system. 

When $G n_\nu T >>1$ we find complete homogenization of flavors and their spectra. Choosing
$G=3\times 10^{-6} (MeV)^{-2}$ as in the model of ref. \cite{ful}, and at the mean neutrino number densities in the decoupling era, $n_\nu \approx 1. ~MeV^3$ we find that the time needed for homogenization is of the order of $T=10^{-5}$ sec.

\section{6. Moving towards the application.}

Our examples appear to be a long way from addressing a physical application, especially the one sought in ref. \cite{ful}. The latter addresses the finer details of the neutrino distributions that emerge from the decoupling region at a temperature of 1 MeV and a little less. In the history of this system there are never large flavor imbalances, and it appears at first sight that this fact should increase the time scales for the final bit of equilibration, through our mechanism, by some orders of magnitude. The problem that we have been spending our time solving, with successive improvements in order to improve the realism, has remained one in which we begin with the maximum amount of flavor dis-equilibrium, all $\nu_e$ or all $\nu_x$.

For example, if we think about beginning in an exactly particle-antiparticle symmetric medium (zero neutrino chemical potential) \underline{and}  with the number of $\nu_e$'s almost equal to the number of $\nu_x$'s , the first thing we might say is:  ``Now almost nothing can happen. The $\nu_e +\nu_e \leftrightarrow  \bar \nu_x + \bar \nu_x $  mechanism does not act to change this state hardly at all, since at the same time we had an equal number of $ \bar \nu_x$'s transformed back to $\nu_e$'s." That  would be the inevitable conclusion of a formulation that is based solely on the usually introduced variables, once the seedings are in hand, that is to say, on expectations of the Fermi field products as they are conventionally enumerated in standard references, e.g. ref. \cite{rs}.

To understand why this reasoning does not apply to the case at hand, consider the momentum states that enter a calculation. For clarity, let us start at t=0 in a system in which we have nearly equal or equal numbers, $2 (N_1+N_2)$ of left-moving (L ) momenta and $2(N_1+N_2)$ of right-moving momenta (R), all with respect to some x-axis. The detailed momentum sets are obtained by picking from the Fermi distribution with zero chemical potential. Then we randomly assign $2 N_1$ of the momentum states to the flavor $\nu_e$  and $2 N_2$ to the flavor $\bar \nu_x$.
We end with four groups, in our initial state $|ABCD \rangle $,
respectively given by,\newline
\newline
A. $\nu_e [L]$, ${\bf q_1,q_2....q_{N_1}}$  ~~~~~~~~          B. $ \nu_e[R]$, ${\bf p_1,p_2,...p_{ N_1}}$ \newline
\newline
C. $\bar\nu_x[L]$, ${\bf r_1,r_2....r_ {N_2}}$ ~~~~~~~~       D. $\bar\nu_x[R]$, ${\bf s_1,s_2...s_ {N_2}}$\newline
\newline

We keep in mind that there is an equally big set of states that were initially just the anti-particles of the above (but with different momenta draws). The latter do not affect the evolution of the former, however.
Now we mainly have the reactions R+L$\rightarrow$R+L, rather than R+R$\rightarrow$R+R or
L+L$\rightarrow$L+L.  In the simplified argument below we shall use only R+L interactions.

The initial momentum distributions in each set are derived from Fermi distributions with zero chemical potential (for particle-antiparticle symmetry). Then, independently of the temperature (as long as it is relativistic), we have a probability 
$(- \pi^2/9+ \zeta[3])/ \zeta[3])<$.1
 that the momentum of a randomly drawn state, say from group A matches  \underline{any} of the momenta in group C, or drawn from B that matches D. 
These sets are thus nearly disjoint, for typical throws of the dice, and the small minority of overlapping states can be written off as duds.

Thus the calculation of progress of the system as it develops from an 
initial $|\rm {ABCD} \rangle$ wave function in the $2(N_1+N_2) $dimensional space is the same as that which can be calculated from separate calculations in $2N_1$ or $2N_2$ dimensional spaces with initial respective wave functions $|\rm{AB}\rangle$ and 
$|\rm{CD}\rangle$, then taking the direct product. That is, they can be calculated precisely 
from initial configurations with maximum flavor imbalance.
Now suppose for example that the initial electron neutrinos have a slightly higher temperature than the x-neutrinos, or a slightly greater number density. In the above description that disequilibrium would be addressed  in a very small number of turnover times. We shall choose coupling constants, $G$, for the effective scalar four-Fermi coupling quite a bit smaller than used in ref.\cite{ful}, but still greater than $G_F$.

\section{7. Adding mixing with a sterile neutrino, $\nu_s$.} 
As an application of the structures that we have been developing, we add an ordinary $\nu_e$-$\nu_s$ mixing term to the system.  We revert to
the head-on clashing ``mono-energetic"  beam model for the active neutrino clouds, in this first pass at a new problem.
The single particle annihilation operators, as in sec.1, were $a,c$ for the 
first beam and $b,d$, leading to the quadratics $\vec \sigma$ and $\vec\tau$, expressed as 
$\sigma_{\pm}, \sigma_3$. To add mixings of, say, $\nu_e$ with a sterile $\nu_s$ we introduce the annihilator $s$ for a sterile $\nu_s$ in the R directional beam and introduce four new operators in which it enters as a factor. The whole set is now,

\begin{eqnarray}
&\sigma_+=c^\dagger a~~~~,~~~~\tau_+=d^\dagger b~,
\nonumber\\
&\sigma_3=a^\dagger a-c^\dagger c~~~,~~~~\tau_3=b^\dagger b-d^\dagger d~,
\nonumber\\
&\zeta_+=a^\dagger s~,~\zeta_3=a^\dagger a-s^\dagger s ~,
\nonumber\\
&~\rho_+=c^\dagger s ~~~~,~~~~ \eta=s^\dagger s~.
\nonumber\\
\,
\end{eqnarray}

The Hamiltonian is given by (\ref{ham1}) with an added active-sterile mixing term,
\begin{eqnarray}
&H_{\rm eff} = { 4\pi G  \over   V} \Bigr [ \sigma_+  \tau_++ \sigma_- \tau_-\Bigr ]
 \nonumber\\
& +b\sin \theta (\zeta_+ +\zeta_ - )+b \eta]~,
\nonumber\\
\,
\label{hams}
 \end{eqnarray}
 where $\theta$ is the mixing angle for $\nu_e$ and $\nu_s$ mixing and we have defined 
 $b=m_s^2 / 2 E$.  For simplicity, we shall take no direct mixing between $\nu_x$ and $\nu_s$. The eight coupled equations are, 
 \begin{eqnarray}
&i\dot \sigma_+=  g \,\tau_-\, \sigma_3 -b \sin \theta \rho_+ ~~~~,~~~i \dot \tau_+= g\, \sigma_-\tau_3\,,
\nonumber\\
&i\dot\sigma_3= 2  g(\sigma_+\, \tau_+ -\sigma_- \,\tau_-)\,+b \sin \theta ~(\zeta_+-\zeta_-) \,,
\nonumber\\
&i\dot\tau_3= 2  g(\sigma_+\, \tau_+ -\sigma_- \,\tau_-)\, ,
\nonumber\\
&i\dot \rho_+=- g \zeta_+ \tau_+ +b \sin \theta ~\sigma_+ + b \rho_+\,,
\nonumber\\
&i\dot \zeta_+=  -g \tau_- \rho_+ +b \sin \theta~\zeta_3 +b\, \zeta_+\,,
\nonumber\\
&i\dot \zeta_3=  g(\sigma_+\, \tau_+ -\sigma_- \,\tau_-)\,-b \sin \theta~(\zeta_+-\zeta_-) \,,
\nonumber\\
&i\dot \eta=-b \sin \theta~ (\zeta_+ -\zeta_-)\,.
\label{eom6}
\end{eqnarray}

We follow the basic mean field approach, in which the initial value of each one of the variables is replaced by its expectation value. At the present time we are not able to follow the modified mean-field approach because of computer limitations and a much larger equation set. Earlier we discussed the alternative
of seeding the initial values, $\sigma_\pm (t=0)=\pm i/N$,  and similarly for $\tau_\pm (t=0)$, which should give approximate results.  In our example, for a pure $\nu_e$ gas, 
we have $\sigma_3(t=0)=1$, $\tau_3(t=0)=1$ after the rescalings used earlier.

For now the point of the equations \ref{eom6} is to show that there really is a way in which the breaks as plotted in fig. 3 , when supplemented with a standard active-sterile mixing term, may be able to 
provide a new time dependent mechanism for shaking the steriles loose. The argument will be on the delicate side since the basic turnover curves as in fig. 3

The solid curve in fig. 4 again shows another example of a solution to the system for an initial state that is all
$\nu_e$ in both directions, on its way to becoming a state with all $\bar \nu_x$ as in the curves of fig. 3,
now with the variable $\sigma_3$ measuring this progress. The dashed curve is produced by adding
the active-sterile interaction, (\ref{hams}), in which only $\nu_e$ couples to the sterile
$\nu_s$, and $\bar \nu_e $ to $\bar \nu_s$. Parameters are very roughly set for the temperature range 1 MeV - 3 MeV. Sterile mass $m_s=10$ eV  and mixing angle $\sin\theta_{s,e}=10^{-5}$. The time scale is such that $t=1$ corresponds to 10$^{-7}$ cm.,
The coupling constant of the active $\nu$'s to the scalar meson was taken as  $G=10^{-5}[{\rm MeV}]^{-2}$ as in \cite{ful}.
The curves show active-sterile mixing that clearly has been stimulated by
the $\nu_e ,\nu_x$ mixing that was the focus of the body of this paper. \begin{figure}[h] 
 \centering
\includegraphics[width=2.5 in]{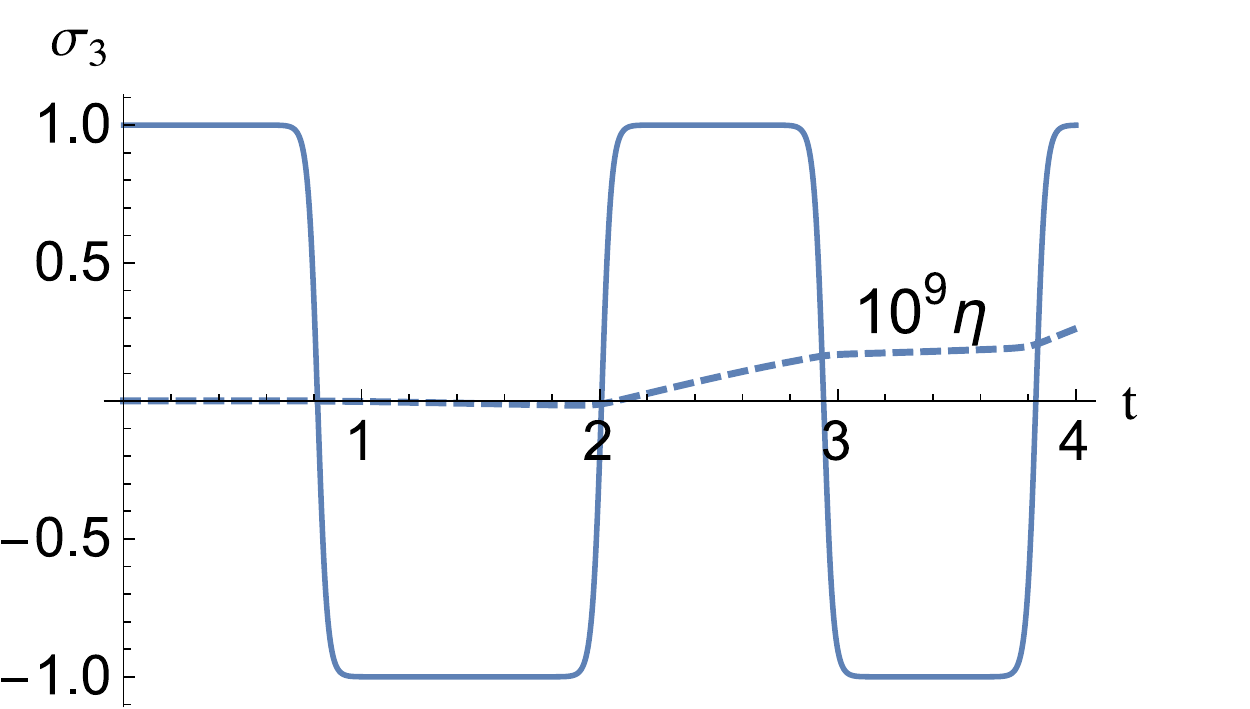}
 \caption{ \small }Solid curve: 0scillation between $\nu_e$ and $\bar \nu_x$  (or $\bar \nu_e$ and $\nu_x$)  as measured by the expectation of $\sigma_3$. Dashed curve: the probability of production of a sterile in any particular mode, as measured by $\eta$ in (\ref{eom6}). 
 \end{figure}

The conventional description of sterile neutrino production for the small mixing angle case involves 
repeated neutrino collisions, each one of which adds a bit of sterile, with tiny probability $\sin^2 \theta_{s,e}$.
We have proposed instead a mechanics whereby the individual active $\nu$'s may be converted to $\nu_s$, with about the same probability, but with a time-scale that is shorter by a factor of $10^{10}$ or so than the ordinary collision time for neutrinos in the medium. A prerequisite to make this possible is to have the instability at the mean-field level. But we also need to know more about the initiation and extinction of the unstable behavior. We favor a picture in which each local breakout of an unstable oscillation lasts a few oscillations at most and then fades, but in which such things are at all times being created in every direction everywhere in our medium.
It is interesting but probably premature to  speculate that this could provide a fast way of populating the universe with dark matter by producing mass =1 KeV sterile neutrinos during the conventional neutrino decoupling era. We must go very carefully in view of the fact that a 3 MeV beam maintains it's phase coherence (in the contexts where this is needed) for only about 10 meters.

 \section{8. Conclusion}

In clouds of massless particles that interact with each other, and in which the particles have some set of flavors (e,x,$\bar{\rm e},\bar{\rm x}$ in the present neutrino case) we have a multiplicity of two body reaction channels in which flavors change, or which momentum goes with which flavor changes, while the set of momenta of the whole swarm stays the same. Then one may find an instability leading to exponential growth of perturbations, and a time-scale for total states-mixing  of order $G^{-1}$, where $G$ is the effective four-Fermi coupling induced by the coupling to the introduced scalar meson. We have here examined its possible role in neutrino flavor equilibration in the neutrino-decoupling era of the early universe. We find possible impact both on the thermal history and on a speculative model of dark matter production.

Of course, actual neutrinos have mass, and the above will be limited to distances $D$ less than
$(m_\nu^2/|\vec p|)^{-1}$, in order to maintain coherence. Followers of the neutrino physics calculations
in the type-2 supernovae calculations will remember a ``fast transformation" possibility for that system as well, where the coupling was through Z meson exchanges. But that demanded rather specific attributes of the initial flavor and angular distributions. In contrast, in the scalar coupling model used in ref \cite{ful} and in the present paper, the instabilities are ubiquitous. We expect them to
dominate, as long as the scalar coupling constant is greater than $G_F$.

We return to one question posed in the introduction, ``What would the effects on the outgoing neutrino
spectra and flavor-balance be, given our treatment of the $\nu+\nu \leftrightarrow \bar\nu+ \bar \nu $ 
process as described in \cite{ful}, but with no account of the back reaction on other physics of the era." The answer
is: complete equilibration of flavors and energies in the entire range $10^{-10}<G^2<3 \times10^{-6}$MeV$^{-2}$. For values moderately less than $10^{-10}$ it could still do the job, depending on how it coexists with the purely refractive part of the standard model interactions. That is to say: the transformations would still be very fast compared to the scattering rate.

The issues raised in sections 1-5, should be further explored with respect to all of the conclusions of ref. \cite{deg},\cite{ful} and other work on the effects of $\nu+\nu\leftrightarrow \bar\nu +\bar\nu$ in the decoupling era. The results of the sec. 6 discussion bearing on efficient production of sterile neutrinos are still a bit provisional. We would love to have been able to solve the generalization of equation set (\ref{eoma}) for cases with many beams, to show that isotropy is not an issue here either, but we do not have the computational power.

\end{document}